%% file: main.tex
\pdfoutput=1

\documentclass[12pt,a4paper]{article}

\usepackage{ifthen} 
\newboolean{pdflatex}
\setboolean{pdflatex}{true} 

\newboolean{articletitles}
\setboolean{articletitles}{true} 

\newboolean{uprightparticles}
\setboolean{uprightparticles}{false} 

\usepackage{amssymb}
\usepackage{amsfonts}
\usepackage{upgreek} 
\usepackage{subfig}
\usepackage{pdflscape}
\newsubfloat{figure}
\newsubfloat{table}

\usepackage[abs]{overpic}
\usepackage{verbatim}

\input{preamble}
\begin{document}

\renewcommand{\thefootnote}{\fnsymbol{footnote}}
\setcounter{footnote}{1}

\input{title-LHCb-PAPER}

\renewcommand{\thefootnote}{\arabic{footnote}}
\setcounter{footnote}{0}



\pagestyle{plain} 
\setcounter{page}{1}
\pagenumbering{arabic}

\linenumbers

%

\input{introduction}

\input{detector}

\input{selection}

\input{likelihoods}

\input{normalization}

\input{backgrounds}

\input{results}

\input{acknowledgements}

\addcontentsline{toc}{section}{References}
\bibliographystyle{LHCb}
\bibliography{main}

\end{document}

%% file: preamble.tex
\usepackage{microtype}
\usepackage{lineno}  
\usepackage{xspace} 

\usepackage{graphicx}  
\usepackage{color}
\usepackage{colortbl}
\graphicspath{{./figs/}} 

\usepackage{amsmath} 
\usepackage{amssymb}
\usepackage{amsfonts}
\usepackage{upgreek} 

\newcommand*\patchAmsMathEnvironmentForLineno[1]{%
\expandafter\let\csname old#1\expandafter\endcsname\csname #1\endcsname
\expandafter\let\csname oldend#1\expandafter\endcsname\csname
end#1\endcsname
 \renewenvironment{#1}%
   {\linenomath\csname old#1\endcsname}%
   {\csname oldend#1\endcsname\endlinenomath}%
}
\newcommand*\patchBothAmsMathEnvironmentsForLineno[1]{%
  \patchAmsMathEnvironmentForLineno{#1}%
  \patchAmsMathEnvironmentForLineno{#1*}%
}
\AtBeginDocument{%
\patchBothAmsMathEnvironmentsForLineno{equation}%
\patchBothAmsMathEnvironmentsForLineno{align}%
\patchBothAmsMathEnvironmentsForLineno{flalign}%
\patchBothAmsMathEnvironmentsForLineno{alignat}%
\patchBothAmsMathEnvironmentsForLineno{gather}%
\patchBothAmsMathEnvironmentsForLineno{multline}%
}

\usepackage{hyperref}    
\usepackage[all]{hypcap} 

\input{lhcb-symbols-def} 
\input{our-symbols-def} 
\usepackage{cite} 
\usepackage{mciteplus}

%% file: lhcb-symbols-def.tex
\def\lhcb {\mbox{LHCb}\xspace}
\def\ux85 {\mbox{UX85}\xspace}

\def\babar  {\mbox{BaBar}\xspace}
\def\belle  {\mbox{Belle}\xspace}

\ifthenelse{\boolean{uprightparticles}}%
{

 \def\Ppi         {\ensuremath{\uppi}\xspace}

 \def\Ptau        {\ensuremath{\uptau}\xspace}

 \def\PDelta      {\ensuremath{\Delta}\xspace}                 
 \def\PXi      {\ensuremath{\Xi}\xspace}                 
 \def\PLambda      {\ensuremath{\Lambda}\xspace}                 
 \def\PSigma      {\ensuremath{\Sigma}\xspace}                 
 \def\POmega      {\ensuremath{\Omega}\xspace}                 
 \def\PUpsilon      {\ensuremath{\Upsilon}\xspace}                 
 
 \def\PB      {\ensuremath{\mathrm{B}}\xspace}                 
                  
 \def\PD      {\ensuremath{\mathrm{D}}\xspace}

 \def\PK      {\ensuremath{\mathrm{K}}\xspace}

 \def\Pb      {\ensuremath{\mathrm{b}}\xspace}                 
 \def\Pc      {\ensuremath{\mathrm{c}}\xspace}

 \def\Pi      {\ensuremath{\mathrm{i}}\xspace}

 \def\Pp      {\ensuremath{\mathrm{p}}\xspace}

 \def\Ps      {\ensuremath{\mathrm{s}}\xspace}

}
{

 \def\Ppi         {\ensuremath{\pi}\xspace}

 \def\Ptau        {\ensuremath{\tau}\xspace}

 \mathchardef\PDelta="7101
 \mathchardef\PXi="7104
 \mathchardef\PLambda="7103
 \mathchardef\PSigma="7106
 \mathchardef\POmega="710A
 \mathchardef\PUpsilon="7107
                  
 \def\PB      {\ensuremath{B}\xspace}                 
                  
 \def\PD      {\ensuremath{D}\xspace}

 \def\PK      {\ensuremath{K}\xspace}

 \def\Pb      {\ensuremath{b}\xspace}                 
 \def\Pc      {\ensuremath{c}\xspace}

 \def\Pi      {\ensuremath{i}\xspace}

 \def\Pp      {\ensuremath{p}\xspace}

 \def\Ps      {\ensuremath{s}\xspace}

}

\def\squark    {\ensuremath{\Ps}\xspace}

\def\cquark    {\ensuremath{\Pc}\xspace}

\def\bquark    {\ensuremath{\Pb}\xspace}

\def\kaon  {\ensuremath{\PK}\xspace}
  \def\Kbar  {\kern 0.2em\overline{\kern -0.2em \PK}{}\xspace}

\def\Kz    {\ensuremath{\kaon^0}\xspace}
\def\Kzb   {\ensuremath{\Kbar^0}\xspace}
\def\KzKzb {\ensuremath{\Kz \kern -0.16em \Kzb}\xspace}
\def\Kp    {\ensuremath{\kaon^+}\xspace}
\def\Km    {\ensuremath{\kaon^-}\xspace}

\def\KpKm  {\ensuremath{\Kp \kern -0.16em \Km}\xspace}

  \def\Dbar    {\kern 0.2em\overline{\kern -0.2em \PD}{}\xspace}
\def\D       {\ensuremath{\PD}\xspace}

\def\Dz      {\ensuremath{\D^0}\xspace}
\def\Dzb     {\ensuremath{\Dbar^0}\xspace}
\def\DzDzb   {\ensuremath{\Dz {\kern -0.16em \Dzb}}\xspace}
\def\Dp      {\ensuremath{\D^+}\xspace}
\def\Dm      {\ensuremath{\D^-}\xspace}

\def\DpDm    {\ensuremath{\Dp {\kern -0.16em \Dm}}\xspace}

\def\B       {\ensuremath{\PB}\xspace}
  \def\Bbar    {\kern 0.18em\overline{\kern -0.18em \PB}{}\xspace}

\def\Bd      {\ensuremath{\B^0}\xspace}
\def\Bs      {\ensuremath{\B^0_\squark}\xspace}

  \def\Y#1S{\ensuremath{\PUpsilon{(#1S)}}\xspace}

\def\Lbar {\ensuremath{\kern 0.1em\overline{\kern -0.1em\PLambda}}\xspace}

\def\BF         {{\ensuremath{\cal B}\xspace}}

\def\BR         {\BF}
\def\ra                 {\ensuremath{\rightarrow}\xspace}
\def\to                 {\ensuremath{\rightarrow}\xspace}

\def\AT#1     {\ensuremath{A_{\mathrm{T}}^{#1}}\xspace}           

\def\C#1      {\ensuremath{\mathcal{C}_{#1}}\xspace}                       
\def\Cp#1     {\ensuremath{\mathcal{C}_{#1}^{'}}\xspace}                    
\def\Ceff#1   {\ensuremath{\mathcal{C}_{#1}^{\mathrm{(eff)}}}\xspace}        
\def\Cpeff#1  {\ensuremath{\mathcal{C}_{#1}^{'\mathrm{(eff)}}}\xspace}       
\def\Ope#1    {\ensuremath{\mathcal{O}_{#1}}\xspace}                       
\def\Opep#1   {\ensuremath{\mathcal{O}_{#1}^{'}}\xspace}                    

\newcommand{\tev}{\ensuremath{\mathrm{\,Te\kern -0.1em V}}\xspace}
\newcommand{\gev}{\ensuremath{\mathrm{\,Ge\kern -0.1em V}}\xspace}
\newcommand{\mev}{\ensuremath{\mathrm{\,Me\kern -0.1em V}}\xspace}
\newcommand{\kev}{\ensuremath{\mathrm{\,ke\kern -0.1em V}}\xspace}
\newcommand{\ev}{\ensuremath{\mathrm{\,e\kern -0.1em V}}\xspace}
\newcommand{\gevc}{\ensuremath{{\mathrm{\,Ge\kern -0.1em V\!/}c}}\xspace}
\newcommand{\mevc}{\ensuremath{{\mathrm{\,Me\kern -0.1em V\!/}c}}\xspace}
\newcommand{\gevcc}{\ensuremath{{\mathrm{\,Ge\kern -0.1em V\!/}c^2}}\xspace}
\newcommand{\gevgevcccc}{\ensuremath{{\mathrm{\,Ge\kern -0.1em V^2\!/}c^4}}\xspace}
\newcommand{\mevcc}{\ensuremath{{\mathrm{\,Me\kern -0.1em V\!/}c^2}}\xspace}

\def\mum  {\ensuremath{\,\upmu\rm m}\xspace}

\def\invfb   {\ensuremath{\mbox{\,fb}^{-1}}\xspace}

\newcommand{\chisq}{\ensuremath{\chi^2}\xspace}

\def\gsim{{~\raise.15em\hbox{$>$}\kern-.85em
          \lower.35em\hbox{$\sim$}~}\xspace}
\def\lsim{{~\raise.15em\hbox{$<$}\kern-.85em
          \lower.35em\hbox{$\sim$}~}\xspace}

\def\sqs   {\ensuremath{\protect\sqrt{s}}\xspace}

\def\pt         {\mbox{$p_{\rm T}$}\xspace}

\def\evtgen     {\mbox{\textsc{EvtGen}}\xspace}
\def\pythia     {\mbox{\textsc{Pythia}}\xspace}

\def\geant      {\mbox{\textsc{Geant4}}\xspace}

\def\photos     {\mbox{\textsc{Photos}}\xspace}

\def\tell1  {TELL1\xspace}
\def\ukl1   {UKL1\xspace}



%% file: our-symbols-def.tex
\newcommand{\CLs}{\ensuremath{\textrm{CL}_{\textrm{s}}}\xspace}

\newcommand{\Bsmumu}{\ensuremath{\Bs\to\mu^+\mu^-}\xspace}
\newcommand{\Bdmumu}{\ensuremath{\Bd\to\mu^+\mu^-}\xspace}

\newcommand{\BuJpsiK}{\ensuremath{B^-\to J/\psi K^-}\xspace}

\newcommand{\Jpsimumu}{\ensuremath{J/\psi\to \mu^+\mu^-}\xspace}
\newcommand{\Jpsi}{\ensuremath{J/\psi}\xspace}

\newcommand{\tmmm}{\ensuremath{\tau^-\to \mu^-\mu^+\mu^-}\xspace}
\newcommand{\tpmmOS}{\ensuremath{\tau^-\to \bar{p}\mu^+\mu^-}\xspace}
\newcommand{\tpmmSS}{\ensuremath{\tau^-\to p\mu^-\mu^-}\xspace}
\newcommand{\tpmm}{\ensuremath{\tau\to p\mu\mu}\xspace}

\newcommand{\DsPhiPi}{\ensuremath{D_s^-\to \phi (\mu^+\mu^-) \pi^-}\xspace}
\newcommand{\DsPhiKKPi}{\ensuremath{D_s^-\to \phi (K^+K^-) \pi^-}\xspace}
\newcommand{\DsTauNu}{\ensuremath{D_s^-\to \tau^- \bar{\nu}_{\tau}}\xspace}
 
\newcommand{\Phimm}{\ensuremath{\phi\to \mu^+\mu^-}\xspace} 
\newcommand{\PhiKK}{\ensuremath{\phi\to K^+K^-}\xspace} 
\newcommand{\DsEtaMuNu}{\ensuremath{\D_s^- \to \eta (\mu^+\mu^-\gamma) \mu^- {\bar{\nu}_\mu}\xspace}}

\newcommand{\BRof}[1]{\ensuremath{{\cal B}(#1)}\xspace}
\newcommand{\gl}{\ensuremath{{\rm \mathcal{M}_{3body}}}\xspace}
\newcommand{\pid}{\ensuremath{{\rm \mathcal{M}_{PID}}}\xspace}



%% file: title-LHCb-PAPER.tex

\begin{titlepage}
\pagenumbering{roman}

\vspace*{-1.5cm}
\centerline{\large EUROPEAN ORGANIZATION FOR NUCLEAR RESEARCH (CERN)}
\vspace*{1.5cm}
\hspace*{-0.5cm}
\begin{tabular*}{\linewidth}{lc@{\extracolsep{\fill}}r}
\ifthenelse{\boolean{pdflatex}}
{\vspace*{-2.7cm}\mbox{\!\!\!\includegraphics[width=.14\textwidth]{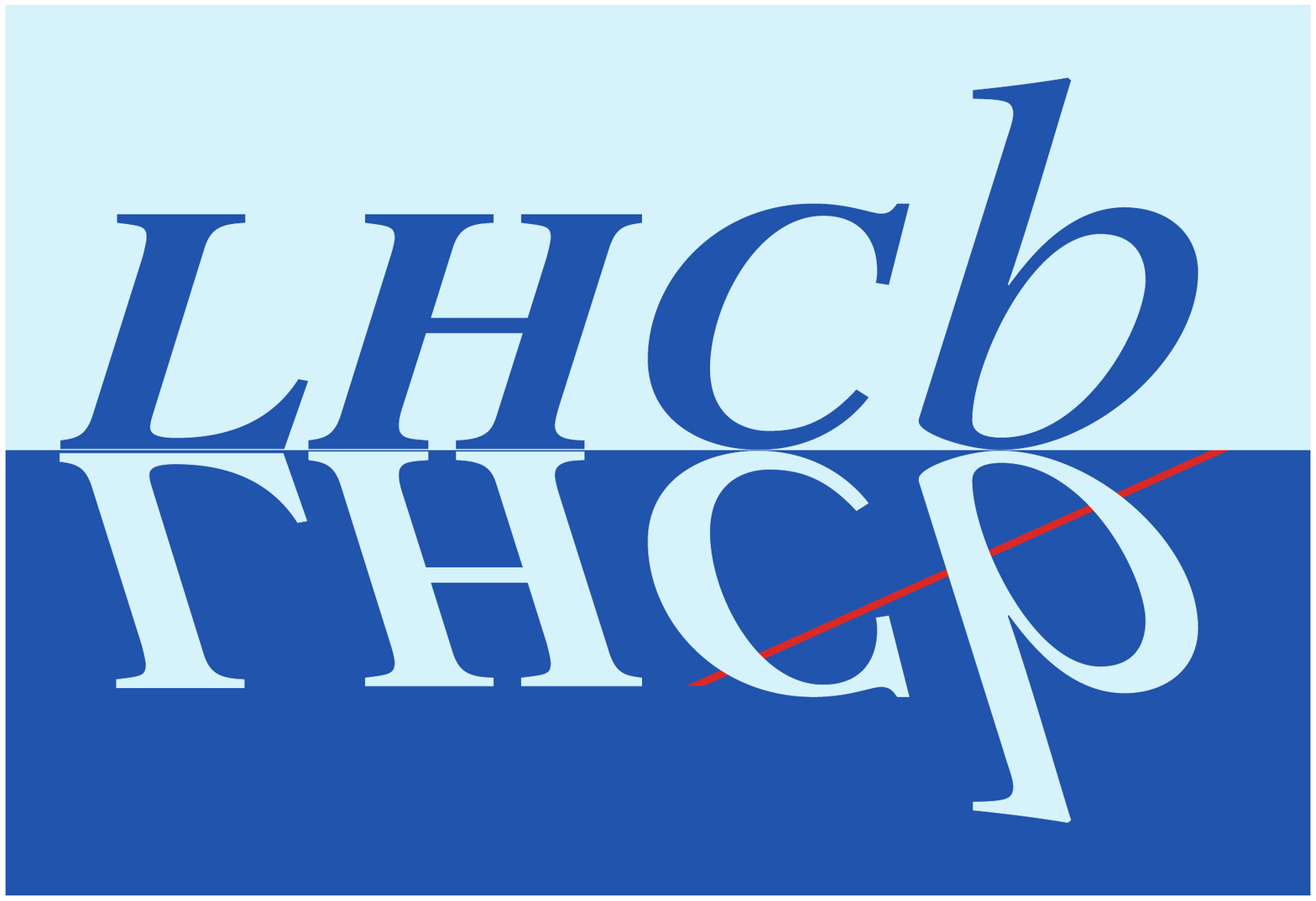}} & &}%
{\vspace*{-1.2cm}\mbox{\!\!\!\includegraphics[width=.12\textwidth]{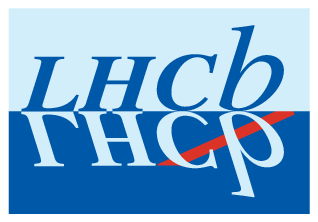}} & &}%
\\
 & & CERN-PH-EP-2013-062 \\  
 & & LHCb-PAPER-2013-014 \\  
 & & \today \\ 
 & & \\
\end{tabular*}

\vspace*{4.0cm}

{\bf\boldmath\huge
\begin{center}
  Searches for violation of lepton flavour and baryon number in tau lepton decays at LHCb
\end{center}
}

\vspace*{2.0cm}

\begin{center}
The LHCb collaboration\footnote{Authors are listed on the following pages.}
\end{center}

\vspace{\fill}

\begin{abstract}
  \noindent
Searches for the lepton flavour violating decay \tmmm and the lepton flavour
and baryon number violating decays \tpmmOS and \tpmmSS have
been carried out using proton-proton collision data, corresponding to
an integrated luminosity of $1.0$~\invfb, taken by the 
\lhcb experiment at $\sqs = 7 \tev$. No evidence has been found for 
any signal, and limits 
have been set at $90\%$ confidence level on the branching fractions: $\BF(\tmmm) < 8.0 \times 10^{-8}$,
$\BF(\tpmmOS) < 3.3 \times 10^{-7}$ and $\BF(\tpmmSS) < 4.4 \times 10^{-7}$.  
The results for
the \tpmmOS and \tpmmSS decay modes represent the first direct experimental 
limits on these channels.
\end{abstract}

\vspace*{1.0cm}

\begin{center}
  Submitted to Physics Letters B
\end{center}

\vspace{\fill}

{\footnotesize 
\centerline{\copyright~CERN on behalf of the \lhcb collaboration, license \href{http://creativecommons.org/licenses/by/3.0/}{CC-BY-3.0}.}}
\vspace*{2mm}

\end{titlepage}


\newpage
\setcounter{page}{2}
\mbox{~}
\newpage

\input{LHCb_authorlist.tex}

\cleardoublepage

%% file: LHCb_authorlist.tex
\centerline{\large\bf LHCb collaboration}
\begin{flushleft}
\small
R.~Aaij$^{40}$, 
C.~Abellan~Beteta$^{35,n}$, 
B.~Adeva$^{36}$, 
M.~Adinolfi$^{45}$, 
C.~Adrover$^{6}$, 
A.~Affolder$^{51}$, 
Z.~Ajaltouni$^{5}$, 
J.~Albrecht$^{9}$, 
F.~Alessio$^{37}$, 
M.~Alexander$^{50}$
% ...

{\footnotesize \it
% ...
$ ^{5}$Clermont Universit\'{e}, Universit\'{e} Blaise Pascal, CNRS/IN2P3, LPC, Clermont-Ferrand, France\\
$ ^{6}$CPPM, Aix-Marseille Universit\'{e}, CNRS/IN2P3, Marseille, France\\
% ...
$ ^{9}$Fakult\"{a}t Physik, Technische Universit\"{a}t Dortmund, Dortmund, Germany\\
% ...
$ ^{35}$Universitat de Barcelona, Barcelona, Spain\\
$ ^{36}$Universidad de Santiago de Compostela, Santiago de Compostela, Spain\\
$ ^{37}$European Organization for Nuclear Research (CERN), Geneva, Switzerland\\
% ...
$ ^{40}$Nikhef National Institute for Subatomic Physics, Amsterdam, The Netherlands\\
% ...
$ ^{45}$H.H. Wills Physics Laboratory, University of Bristol, Bristol, United Kingdom\\
% ...
$ ^{50}$School of Physics and Astronomy, University of Glasgow, Glasgow, United Kingdom\\
$ ^{51}$Oliver Lodge Laboratory, University of Liverpool, Liverpool, United Kingdom\\
% ...
$ ^{n}$LIFAELS, La Salle, Universitat Ramon Llull, Barcelona, Spain\\
% ...
}
\end{flushleft}

%% file: introduction.tex

\section{Introduction}
\label{sec:Introduction}

The observation of neutrino oscillations was the first
evidence for lepton flavour violation (LFV). As a consequence, the 
introduction of mass terms for neutrinos in the Standard Model (SM)
implies that LFV exists also in the charged
sector, but with branching fractions smaller than 
$\sim 10^{-40}$~\cite{Raidal:2008jk, Ilakovac:2012sh}.
Physics beyond the Standard Model (BSM) could significantly enhance these branching fractions. Many BSM theories
predict enhanced LFV in $\tau^-$ decays with respect to $\mu^-$ decays\footnote{The inclusion of charge conjugate processes is implied throughout this Letter.}, with branching
fractions within experimental reach \cite{lfvreview}. To date, no charged LFV decays 
such as $\mu^-\to e^-\gamma$, $\mu^-\to e^-e^+e^-$,
$\tau^-\to\ell^-\gamma$ and $\tau^-\to\ell^-\ell^+\ell^-$ (with $\ell^-=e^-,\mu^-$)
have been observed~\cite{Amhis:2012bh}. Baryon number violation (BNV) is believed
to have occurred in the early universe, although the mechanism is unknown.
BNV in charged lepton decays automatically
implies lepton number and lepton flavour violation, with angular momentum 
conservation requiring the change $|\Delta( B - L ) | = 0$ or $2$, where $B$ and 
$L$ are the net baryon and lepton numbers.
The SM and most of its extensions~\cite{Raidal:2008jk} require $|\Delta( B - L ) | = 0$.
Any observation of BNV or charged LFV would 
be a clear sign for BSM physics, while a lowering of the experimental upper limits on branching
fractions would further constrain the parameter spaces of BSM models.

In this Letter we report on searches for the LFV decay \tmmm and the LFV and BNV 
decay modes \tpmmOS and \tpmmSS at \lhcb~\cite{Alves:2008zz}. 
The inclusive $\tau^-$ production cross-section at the LHC is relatively large, at about $80\,\upmu$b (approximately $80\%$ of which comes from \DsTauNu), estimated using the
$b\bar{b}$ and $c\bar{c}$ cross-sections measured by
\lhcb~\cite{sigmabbLHCb,sigmaccLHCb} and the inclusive $b\ra\Ptau$ and $c\ra
\Ptau$ branching fractions~\cite{PDG}. 
The \tmmm and \tpmm decay modes\footnote{In the following \tpmm refers to both the \tpmmOS and \tpmmSS channels.} are of particular interest at
\lhcb, since 
muons provide clean signatures in the detector and
the ring-imaging Cherenkov (RICH) detectors give excellent identification of
protons. 

This Letter presents the first results on the \tmmm decay mode
from a hadron collider and demonstrates an experimental sensitivity 
at \lhcb, with data corresponding to an integrated luminosity of $1.0$\invfb, that 
approaches the 
current best experimental upper limit, from 
\belle, $\BF(\tmmm)<2.1 \times 10^{-8}$ at 90\% confidence level (CL)~\cite{Hayasaka:2010np}.
\babar and \belle have searched for BNV \Ptau decays with $|\Delta( B - L ) | = 0$ and
$|\Delta( B - L ) | = 2$ using the modes
$\tau^- \to \PLambda h^-$ and $\bar{\PLambda} h^-$ (with $h^- = \Ppi^-,\PK^-$), and upper limits on branching
fractions of
order $10^{-7}$ were obtained~\cite{Amhis:2012bh}. \babar has also searched for
the $B$ meson decays $B^0 \to \PLambda_c^+ l^-$,  $B^- \to \PLambda l^-$
(both having $|\Delta( B - L ) | = 0$) and $B^- \to \bar{\PLambda} l^-$ ($|\Delta( B - L ) | = 2$),
obtaining upper limits at 90\% CL on branching fractions in the range $(3.2 - 520) \times 10^{-8}$~\cite{BABAR:2011ac}.
The two BNV \Ptau decays presented here, \tpmmOS and \tpmmSS, have $|\Delta( B - L ) | = 0$ but they could have rather different BSM interpretations; 
they have not been studied by any previous experiment.

In this analysis the LHCb data sample from 2011, corresponding to
an integrated luminosity of 1.0\invfb collected at $\sqrt{s}=7\tev$, is used. Selection
criteria
are implemented for the three signal modes, \tmmm, \tpmmOS and \tpmmSS, and for the
calibration and normalisation channel, which is $D_s^-\to\phi\pi^{-}$ followed by $\phi\to\mu^{+}\mu^{-}$, 
referred to in the following as \DsPhiPi.  
These initial, cut-based selections are designed to keep
good efficiency for signal whilst reducing the dataset to a
manageable level. To avoid potential bias, $\mu^-\mu^+\mu^-$ and $p\mu\mu$
candidates with mass
within $\pm 30\mevcc~(\approx3\sigma_m)$ of the $\tau$ mass are initially blinded from the analysis, where $\sigma_m$ denotes the expected mass resolution.
For the $3\mu$ channel, discrimination between potential signal and background 
is performed using a three-dimensional binned distribution in two likelihood variables and
the mass of the $\tau$ candidate. One likelihood variable is based on the three-body decay topology 
and the other on muon identification. For the \tpmm channels, the use of the second likelihood function is replaced by cuts on the 
proton and muon particle identification (PID) variables.
The analysis strategy and limit-setting procedure are similar to those used
for the \lhcb analyses of the \Bsmumu and \Bdmumu channels~\cite{Bsmumu2013, Read_02, *Junk_99}.

%% file: detector.tex
\section{Detector and triggers}
\label{sec:Detector}
 
The \lhcb detector~\cite{Alves:2008zz} is a single-arm forward
spectrometer covering the \mbox{pseudorapidity} range $2<\eta <5$,
designed for the study of particles containing \bquark or \cquark
quarks. The detector includes a high precision tracking system
consisting of a silicon-strip vertex detector surrounding the $pp$
interaction region, a large-area silicon-strip detector located
upstream of a dipole magnet with a bending power of about
$4{\rm\,Tm}$, and three stations of silicon-strip detectors and straw
drift tubes placed downstream. The combined tracking system has
momentum resolution $\Delta p/p$ that varies from 0.4\% at 5\gevc to
0.6\% at 100\gevc, and impact parameter resolution of 20\mum for
tracks with high transverse momentum (\pt). Charged hadrons are identified
using two RICH detectors. Photon, electron and
hadron candidates are identified by a calorimeter system consisting of
scintillating-pad and preshower detectors, an electromagnetic
calorimeter and a hadronic calorimeter. Muons are identified by a
system composed of alternating layers of iron and multiwire
proportional chambers. 

The trigger~\cite{LHCbTrigger} consists of a hardware stage, based
on information from the calorimeter and muon systems, followed by a
software stage that applies a full event reconstruction.
The hardware trigger
selects muons with $\pt>1.48\gevc$.
The software trigger requires a two-, three- or four-track
secondary vertex with a high sum of the \pt of
the tracks and a significant displacement from the primary $pp$
interaction vertices~(PVs). At least one track should have $\pt >
1.7\gevc$ and impact parameter chi-squared (IP $\chisq$),
with respect to the
$pp$ collision vertex, greater than 16. The IP $\chisq$ is defined as the
difference between the $\chisq$ of the PV reconstructed with and
without the track under consideration. A multivariate algorithm is used for
the identification of secondary vertices.
 
For the simulation, $pp$ collisions are generated using
\pythia~6.4~\cite{Sjostrand:2006za} with a specific \lhcb
configuration~\cite{LHCb-PROC-2010-056}.  Particle decays
are described by \evtgen~\cite{Lange:2001uf} in which final-state
radiation is generated using \photos~\cite{Golonka:2005pn}. For
the three signal $\tau$ decay channels, the final-state particles
are distributed according to three-body phase space.
The
interaction of the generated particles with the detector, and its
response, are implemented using the \geant
toolkit~\cite{Allison:2006ve, *Agostinelli:2002hh} as described in
Ref.~\cite{LHCb-PROC-2011-006}.

%% file: selection.tex
\section{Signal candidate selection}
\label{sec:selection}

The signal and normalisation channels have the same topology, 
the signature of which is a vertex displaced from the PV, having three
tracks that are reconstructed to give a mass close to that of the $\tau$ lepton 
(or $D_s$ meson for the normalisation channel). In order to 
discriminate against background, well-reconstructed and well-identified
muon, pion and proton tracks are required, with selections on track quality 
criteria and a
requirement of \pt $> 300$\,\mevc. 
Furthermore, for the \tpmm signal and normalisation channels the muon and 
proton candidates must pass loose PID requirements
and the combined \pt
of the three-track system is required to be greater than $4\gevc$. 
All selected tracks are required to have 
IP $\chisq > 9$.
The fitted three-track vertex has
to be of good quality, with a fit $\chi^{2} < 15$, and the measured decay time, $t$, of
the candidate forming the vertex has to be compatible with that of a 
heavy meson or tau lepton ($ct > 100\mum$). Since the $Q$-values in 
decays of charm mesons to $\tau$ are relatively small, 
poorly reconstructed candidates are removed by a cut on 
the pointing angle between the momentum vector of the three-track system and
the line joining the primary and secondary vertices. In the \tmmm channel, signal 
candidates with a $\mu^+\mu^-$
mass within $\pm 20\mevcc$ of the $\phi$ meson mass are removed, and to eliminate irreducible 
background near the signal region arising from the decay
\DsEtaMuNu, candidates with a $\mu^+\mu^-$ mass combination below
$450\mevcc$ are also rejected (see Section~\ref{sec:backgrounds}). Finally, to remove potential
contamination from pairs of reconstructed tracks that arise from the same
particle, same-sign muon pairs with 
mass lower than 250\mevcc are removed in both the \tmmm and \tpmmSS channels.
The signal regions are defined by $\pm 20\mevcc\ (\approx2\sigma_m)$ windows around the nominal \Ptau mass, but
candidates within wide mass windows, of $\pm$400\mevcc for \tmmm decays and $\pm$250\mevcc for \tpmm decays, 
are kept to allow evaluation of the background contributions in the signal regions.
A mass window of $\pm 20\mevcc$ is also used to define the signal region for the \DsPhiPi channel, 
with the $\mu^+\mu^-$ mass required to be within $\pm20\mevcc$ of the $\phi$ meson mass.

%% file: likelihoods.tex
\section{Signal and background discrimination}
\label{sec:likelihoods}

After the selection each $\tau$ candidate is given a probability to be signal or
background according to the values of several likelihoods. 
For \tmmm three likelihoods are used: a three-body likelihood, \gl, a PID likelihood, \pid, 
and an invariant mass likelihood. 
The likelihood \gl uses the properties 
of the reconstructed \Ptau decay to distinguish displaced three-body decays from $N$-body decays (with $N>3$) 
and combinations of tracks from different vertices. 
Variables used include the vertex quality and its displacement from the 
PV, and the IP and fit $\chi^2$ values of the tracks.
The likelihood \pid quantifies the
compatibility of each of the three particles with the muon
hypothesis using information from the RICH detectors, 
the calorimeters and the muon stations; the value of \pid is taken as the
smallest one of the three muon candidates. 
For \tpmm, the use of \pid is
replaced by cuts on PID quantities. 
The invariant mass likelihood uses the 
reconstructed mass of the \Ptau candidate to help discriminate between signal and background.

For the \gl likelihood a boosted decision tree~\cite{Breiman} is used, with the AdaBoost
algorithm~\cite{AdaBoost}, and is implemented 
via the TMVA~\cite{tmva} toolkit. It is trained using 
signal and background samples, both from simulation, where the
composition of the background is a mixture of
$b\bar{b}\rightarrow\mu\mu X$ and $c\bar{c}\rightarrow\mu\mu X$
according to their relative abundance as measured in data. 
The \pid likelihood uses a neural network, which is also trained on simulated events.
The probability density function shapes are calibrated using the
\DsPhiPi control channel and \Jpsimumu data for the \gl and \pid likelihoods, respectively. 
The shape of the signal mass spectrum is modelled using \DsPhiPi data.   
The \gl response as determined using the training from the \tmmm samples is used also for the \tpmm analyses. 

\begin{figure}[t]
\begin{minipage}[b]{0.5\linewidth}
        \centering
        \begin{overpic}[width=0.95\textwidth]{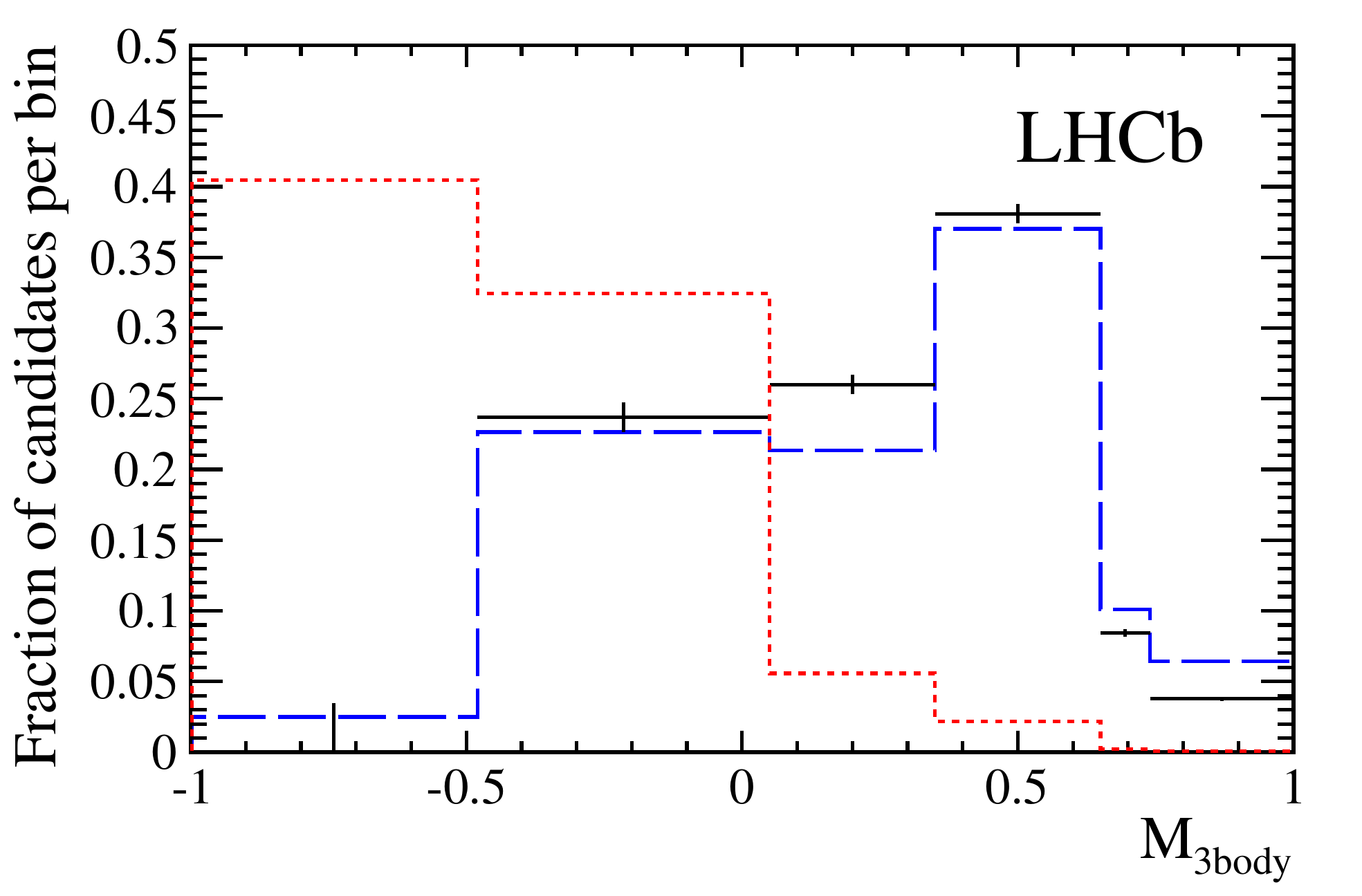}
        \put (40,120) {\small{(a)}}
        \end{overpic}
\end{minipage}
\begin{minipage}[b]{0.5\linewidth}
        \centering
        \begin{overpic}[width=0.95\textwidth]{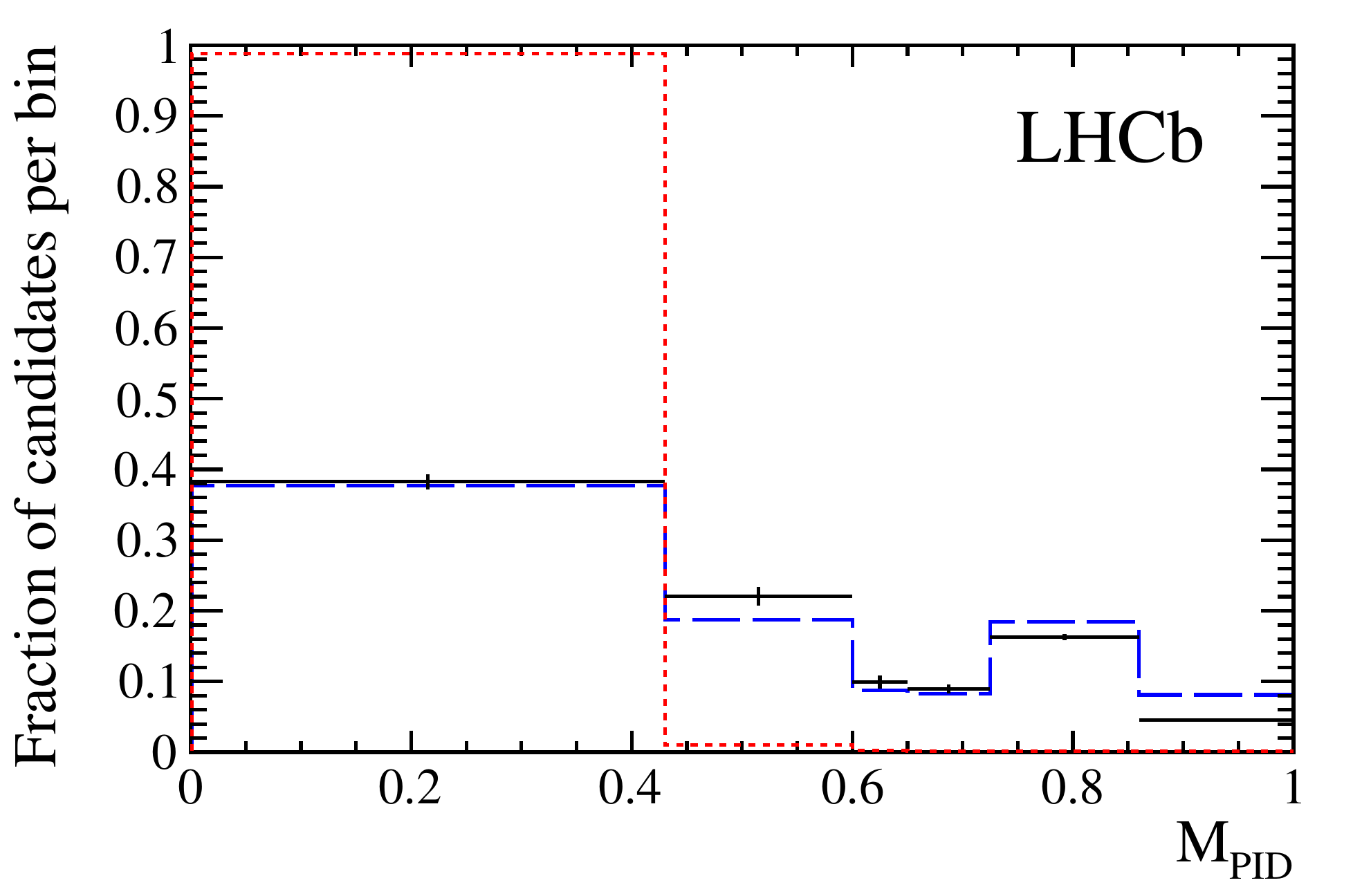}
        \put (40,120) {\small{(b)}}
        \end{overpic}
\end{minipage}
\caption{\small {Distribution of (a) \gl and (b) \pid for \tmmm where the binning corresponds to that used in the limit calculation. 
The short dashed (red) lines show the response of the data sidebands, whilst the long dashed (blue) and solid (black) lines show the response of 
simulated signal events before and after calibration.
Note that in both cases the lowest likelihood bin is later excluded from the analysis.}}
\label{fig:3muPDF}
\end{figure}

For the \gl and \pid likelihoods the binning is chosen such that the separation
power between the background-only and signal-plus-background hypotheses is maximised,
whilst minimising the number of bins. 
For the \gl likelihood the optimum number of bins is found to be six for the \tmmm 
analysis and five for \tpmm, while 
for the \pid likelihood the optimum number of bins is found to be five.
The lowest bins in \gl and \pid do not
contribute to the sensitivity and are later excluded from the analyses.
The distributions of the two likelihoods, along with their binning schemes, are shown in Fig.~\ref{fig:3muPDF} for the \tmmm analysis. 

\begin{figure}[t] 
\begin{minipage}[b]{0.5\linewidth}
	\centering
	\begin{overpic}[width=0.95\textwidth]{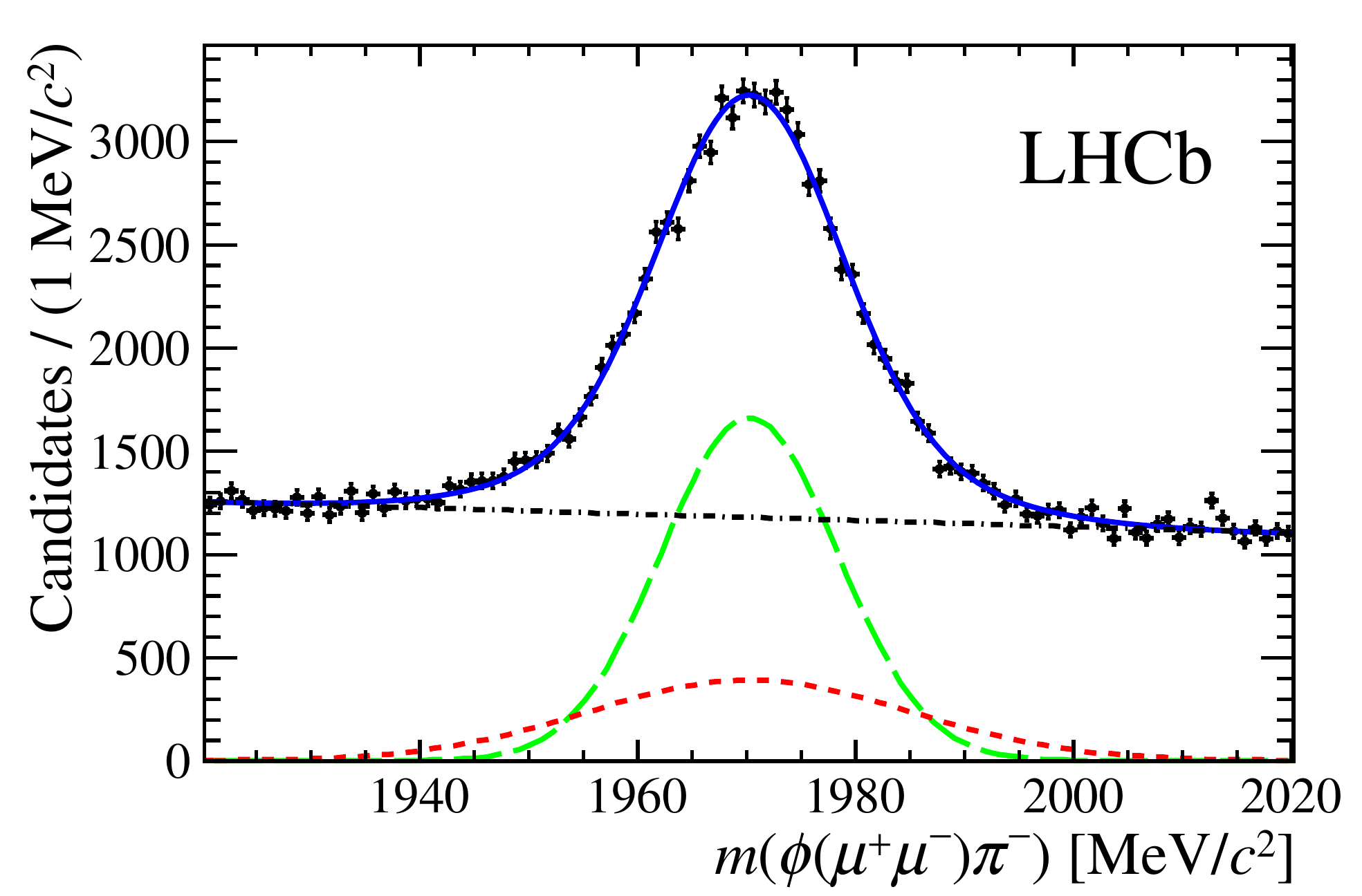}
	\put (40,120) {\small{(a)}}
	\end{overpic}
\end{minipage}
\begin{minipage}[b]{0.5\linewidth}
	\centering
	\begin{overpic}[width=0.95\textwidth]{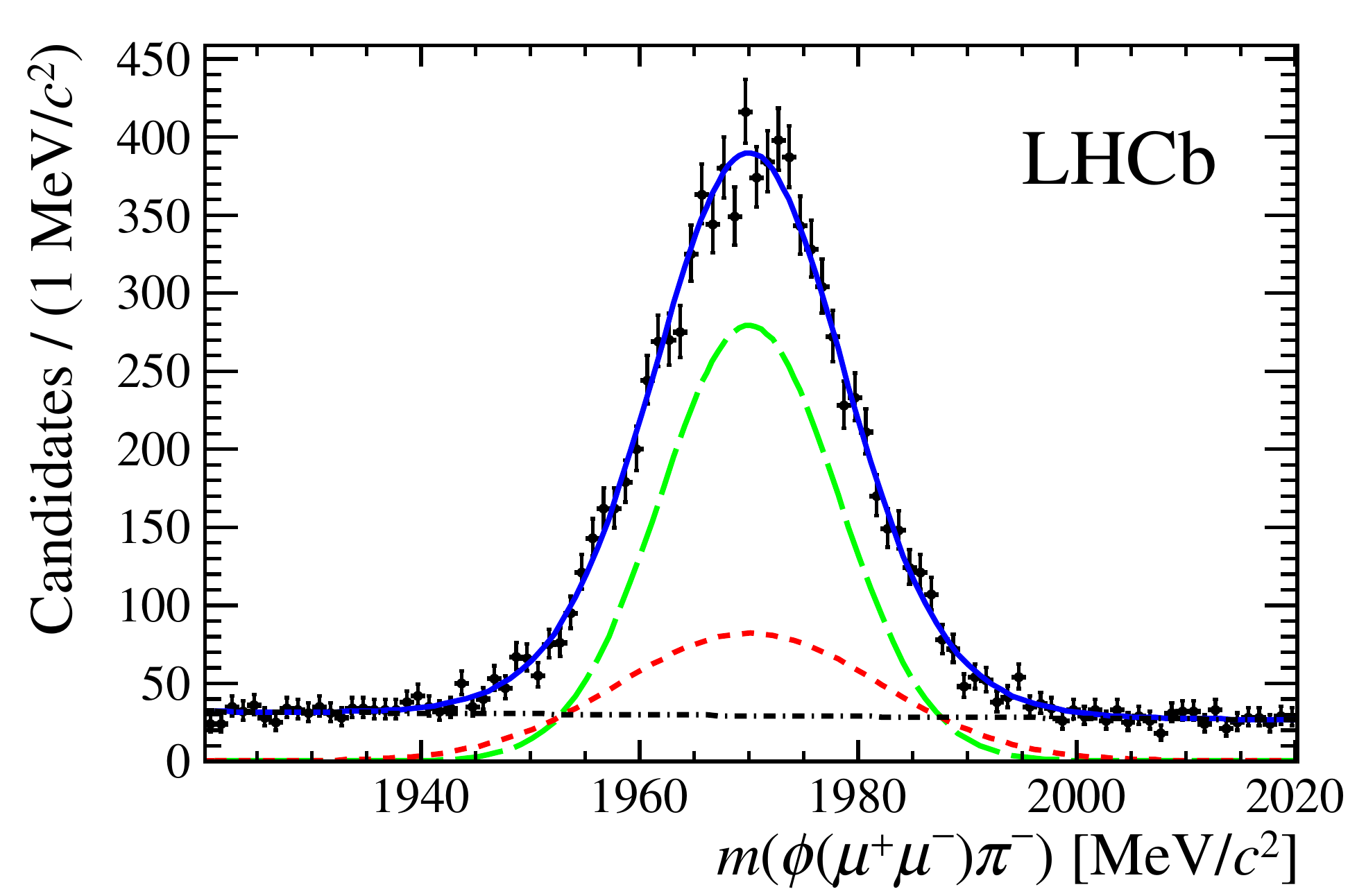}
	\put (40,120) {\small{(b)}}
	\end{overpic}
\end{minipage}
\caption{\small {Invariant mass distribution of $\phi(\mu^+\mu^-)\pi^-$
  after (a) the \tmmm selection and (b) the \tpmm selection and PID cuts. 
The solid (blue) lines show the overall fits, the long dashed (green) and short dashed (red) lines show the two Gaussian 
components of the signal and the dot dashed (black) lines show the backgrounds.}}  
\label{fig:num_Ds}
\end{figure}

For the \tpmm analysis, further cuts on the muon and proton PID hypotheses are used instead of \pid and
are optimised, for a $2\sigma$ significance, on simulated signal events and data sidebands using the figure of merit from Ref.~\cite{punzi}, 
with the distributions of the PID variables corrected according to those observed in data. 
The expected shapes of the invariant mass spectra for the \tmmm and \tpmm signals, with the appropriate selections applied, are 
taken from fits to the \DsPhiPi control channel in data 
as shown in Fig.~\ref{fig:num_Ds}. 
The signal distributions are modelled with 
the sum of two Gaussian functions with a common mean, where the narrower Gaussian   
contributes 70\% of the total signal yield, while the combinatorial backgrounds are modelled 
with linear functions. The expected widths of the $\tau$ signals in data are taken from 
simulation, scaled by the ratio of the widths of the $D_s^-$ peaks in data and simulation. 
The data are divided into eight equally 
spaced bins in the $\pm20\mevcc$ mass window around the nominal $\Ptau$ mass.

%% file: normalization.tex
\section{Normalisation}
\label{sec:normalization}

To measure the signal branching fraction for the decay \tmmm (and similarly for \tpmm) we normalise 
to the \DsPhiPi
calibration channel using 
\begin{align}
&{\BRof\tmmm} \nonumber\\
&\quad = {\BRof\DsPhiPi}
\times
\frac{f^{D_{s}}_{\tau}}{\BRof\DsTauNu}
\times
\frac { \rm {\epsilon\mathstrut_{cal}^{REC\&SEL}}}
{\rm \epsilon\mathstrut_{sig}^{REC\&SEL}}
\times
\frac{\rm {\epsilon\mathstrut_{cal}^{TRIG}}}
{\rm \epsilon\mathstrut_{sig}^{TRIG}}
\times\frac{N_{\rm sig}}{N_{\rm cal}} \nonumber\\
&\quad = \alpha \times N_{\rm sig}\, ,
\label{eq:normalization}
\end{align} 
where $\alpha$ is the overall normalisation factor and $N_{\rm sig}$ is the
number of observed signal events. The branching fraction \BRof\DsTauNu is 
taken from Ref.~\cite{Sheldon2012}.  The quantity $f^{D_{s}}_{\tau}$ is the
fraction of \Ptau leptons that originate from $D_s^-$ decays, calculated using the
$b\bar{b}$ and $c\bar{c}$ cross-sections as measured by
\lhcb~\cite{sigmabbLHCb,sigmaccLHCb} and the inclusive $b\ra\Ptau$, $c\ra
\Ptau$, $b\ra D_{s}$ and $c\ra D_{s}$ branching fractions~\cite{PDG}. 
The corresponding expression for the \tpmm decay is identical
except for the inclusion of a further term,
${\rm \epsilon\mathstrut_{cal}^{PID}}/{\rm \epsilon\mathstrut_{sig}^{PID}}$,
to account for the effect of the PID cuts.

The reconstruction and selection efficiencies, $\rm \epsilon^{REC\&SEL}$, are
products of the detector acceptances for the particular
decays, the muon identification efficiencies and the selection
efficiencies. The combined muon identification and selection efficiency is determined from the
yield of simulated events after the full selections have been
applied. In the sample of simulated events, the track IPs are smeared to describe the 
secondary-vertex resolution of the data.
Furthermore, the events are given weights to adjust the prompt and 
non-prompt $b$ and $c$ particle production fractions to the latest measurements~\cite{PDG}. 
The difference in the result if the weights are varied within their uncertainties is assigned as a systematic uncertainty. 
The ratio of efficiencies is corrected to account for the differences between data and simulation 
in efficiencies
of track reconstruction, muon identification, the $\phi(1020)$ mass window cut in the normalisation channel 
and the $\Ptau$ mass window cut, with all associated systematic uncertainties included.
The removal of candidates in the least sensitive bins in the \gl and \pid classifiers
is also taken into account. 

The trigger efficiency for selected candidates, $\rm \epsilon^{TRIG}$, is
evaluated from simulation while its systematic uncertainty is determined from
the difference between trigger efficiencies of \BuJpsiK decays
measured in data and in simulation.

For the \tpmm channels the PID efficiency for selected and triggered candidates, $\rm \epsilon^{PID}$, is calculated
using data calibration samples of $\Jpsi \rightarrow \mu^{+}\mu^{-}$ and $\PLambda\rightarrow\Pp\pi^{-}$ decays, with the tracks
weighted to match the kinematics of the signal and calibration channels.
A systematic uncertainty of 1\% per corrected final-state track is assigned~\cite{sigmaccLHCb}, as well as a further 1\%
uncertainty to account for differences in the kinematic binning of the calibration samples between the analyses.

The branching fraction of the calibration channel 
is determined from a combination of known branching fractions using
\begin{equation}
\BRof{ \DsPhiPi } = \frac{ \BRof{ \DsPhiKKPi} } 
                           {  \BRof{ \PhiKK } } 
                    \BRof{ \Phimm }=(1.33\pm0.12)\times 10^{-5}\, ,
\end{equation} 
\noindent where \BRof{ \PhiKK } and 
\BRof{ \Phimm } are taken from~\cite{PDG} and $\BRof{ \DsPhiKKPi}$ is taken from the BaBar amplitude 
analysis~\cite{Babar2011a}, which considers only the \PhiKK
resonant part of the $D_s^{-}$ decay. This is
motivated by the negligible contribution of non-resonant $D_s^- \to \mu^+\mu^-\pi^-$
events seen in our data. 
The yields of \DsPhiPi candidates in data,
$N_{\rm cal}$, are determined from the fits to reconstructed $\phi(\mu^+\mu^-)\pi^-$ mass distributions, shown in Fig.~\ref{fig:num_Ds}. 
The variations in the yields if the relative 
contributions of the two Gaussian components are varied in the fits are considered as systematic uncertainties. 
Table~\ref{tab:norm_summary} gives a summary of all contributions to 
$\alpha$; the uncertainties are taken to be uncorrelated.

\clearpage
\renewcommand*\arraystretch{1.5}
\begin{table}[t]
\centering
\caption[]{\small {Terms entering in
  the normalisation factor $\alpha$ for \tmmm, \tpmmOS and \tpmmSS,
 and their combined statistical and systematic uncertainties.}}
\vspace{1mm}
\label{tab:norm_summary}
$
\begin{array}{c|r@{\hspace{1mm}\pm\hspace{1mm}}l|r@{\hspace{1mm}\pm\hspace{1mm}}l|r@{\hspace{1mm}\pm\hspace{1mm}}l}
& \multicolumn{2}{l|}{\tmmm} & \multicolumn{2}{l|}{\tpmmOS} & \multicolumn{2}{l}{\tpmmSS} \\   
\hline
\BRof\DsPhiPi & \multicolumn{3}{r@{\hspace{1mm}\pm\hspace{1mm}}}{(1.33} & \multicolumn{3}{@{}l}{0.12) \times 10^{-5}}\\
\hline
f^{D_{s}}_{\tau} & \multicolumn{3}{r@{\hspace{1mm}\pm\hspace{1mm}}}{0.78} & \multicolumn{3}{@{}l}{0.05}\\
\hline
\BRof\DsTauNu & \multicolumn{3}{r@{\hspace{1mm}\pm\hspace{1mm}}}{0.0561} & \multicolumn{3}{@{}l}{0.0024}\\
\hline
\rm{\epsilon\mathstrut_{cal}}^{REC\&SEL}/
\rm{\epsilon\mathstrut_{sig}}^{REC\&SEL}
& 1.49 & 0.12 & 1.35 & 0.12 & 1.36 & 0.12\\  
\hline
\rm{\epsilon\mathstrut_{cal}}^{TRIG}/
\rm{\epsilon\mathstrut_{sig}}^{TRIG}  
& 0.753 & 0.037 & 1.68 & 0.10 & 2.03 & 0.13\\  
\hline
\rm{\epsilon\mathstrut_{cal}}^{PID}/
\rm{\epsilon\mathstrut_{sig}}^{PID}  
& \multicolumn{2}{l|}{\hspace{10mm}\rm{n/a}} & 1.43 & 0.07 & 1.42 & 0.08\\  
\hline
N_{\rm cal} & 48\,076 & 840 & \multicolumn{4}{c}{8\,145\pm180}\\
\hline 
\alpha & (4.34 & 0.65) \times 10^{-9} & (7.4 & 1.2) \times 10^{-8} & (9.0 & 1.5) \times 10^{-8}\\
\end{array}
$
\end{table}
\renewcommand*\arraystretch{1}

%% file: backgrounds.tex
\section{Background studies}
\label{sec:backgrounds}

The background processes for the decay \tmmm consist mainly of decay chains 
of heavy mesons with three real muons in the final state or with one 
or two real muons in combination with two or one misidentified particles.  
These backgrounds vary smoothly in the mass spectra in the region of the 
signal channel.
The most important peaking background channel is found to be
\DsEtaMuNu, about $80\%$ of which is removed (see Section~\ref{sec:selection}) 
by a cut on the dimuon mass. The small remaining background from this process
is consistent with the smooth variation in the mass spectra of the other backgrounds in the mass range considered in the fit.
Based on simulations, no peaking backgrounds are expected in the \tpmm analyses.  
 
\begin{figure}[h]
\begin{center}
\begin{minipage}[b]{0.48\linewidth}
	\centering
	\begin{overpic}[width=\textwidth]{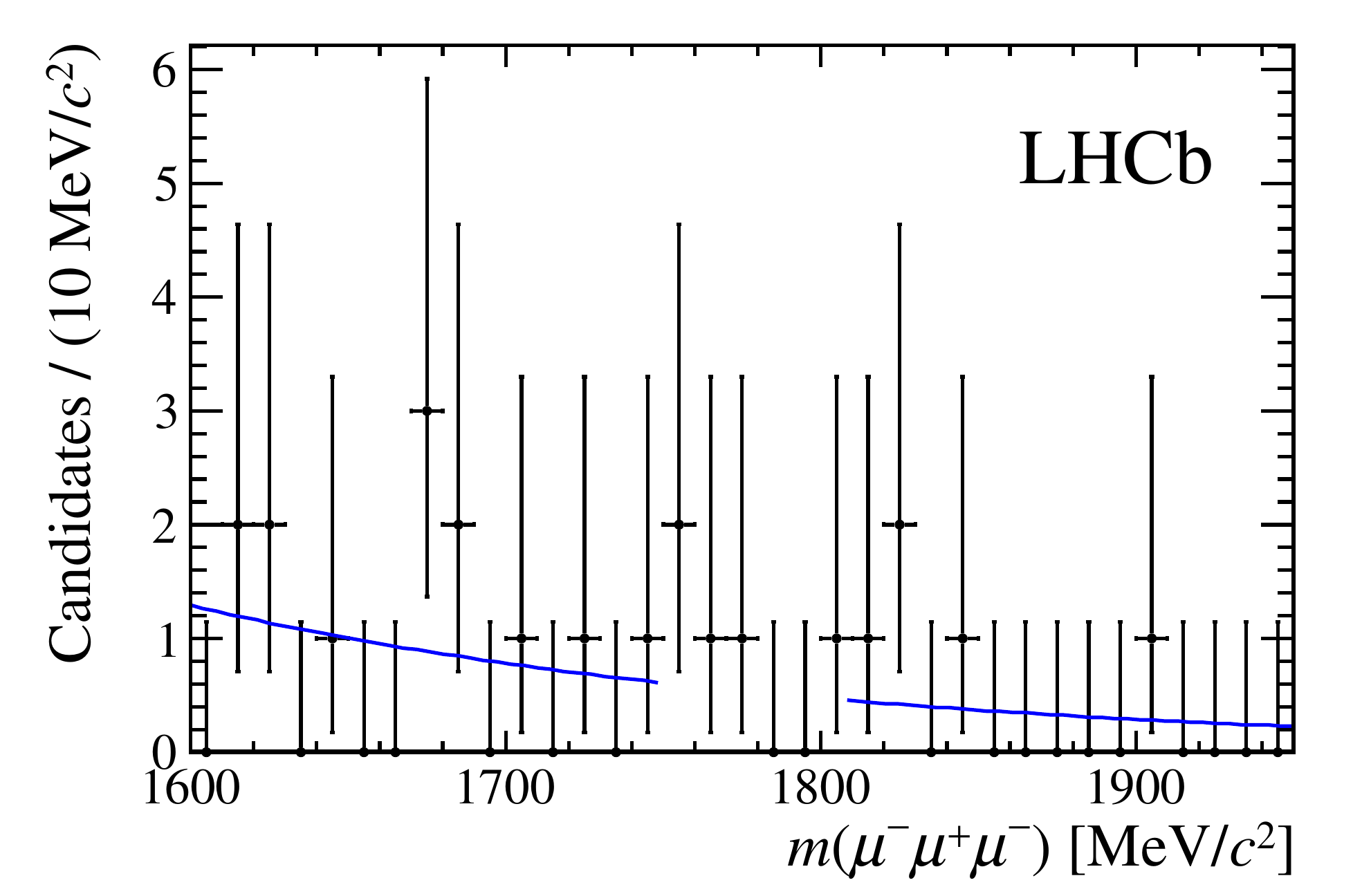}
	\put (40,120) {\small{(a)}}
	\put (80,125) {\tiny{\gl$\in [0.65, 1.0]$}}
	\put (80,115) {\tiny{\pid$\in [0.725, 1.0]$}}
	\end{overpic}
\end{minipage}\\
\begin{minipage}[b]{0.48\linewidth}
	\centering
	\begin{overpic}[width=\textwidth]{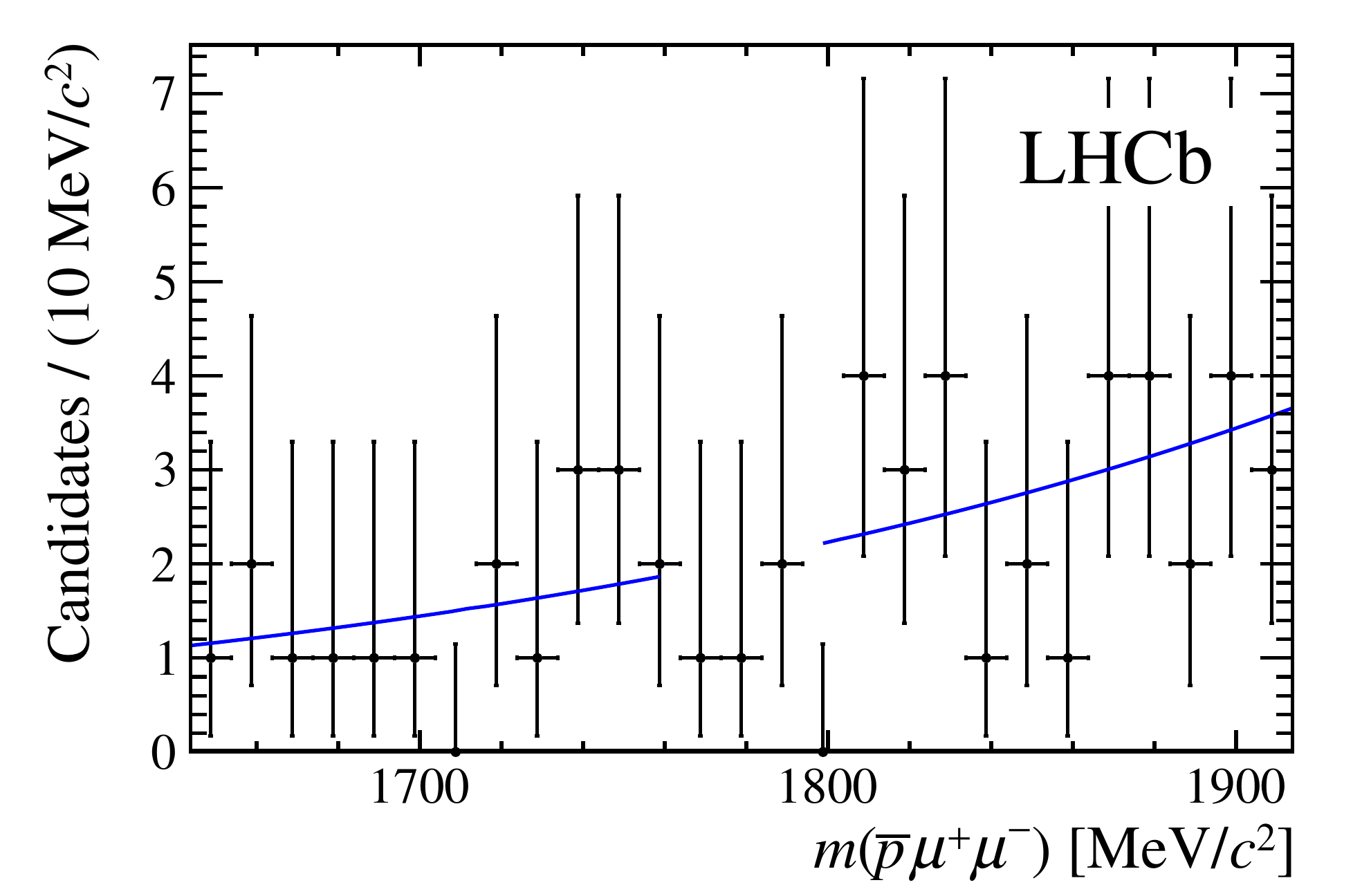}
	\put (40,120) {\small{(b)}}
	\put (60,120) {\tiny{\gl$\in [0.40, 1.0]$}}
	\end{overpic}
\end{minipage}
\begin{minipage}[b]{0.48\linewidth}
	\centering
	\begin{overpic}[width=\textwidth]{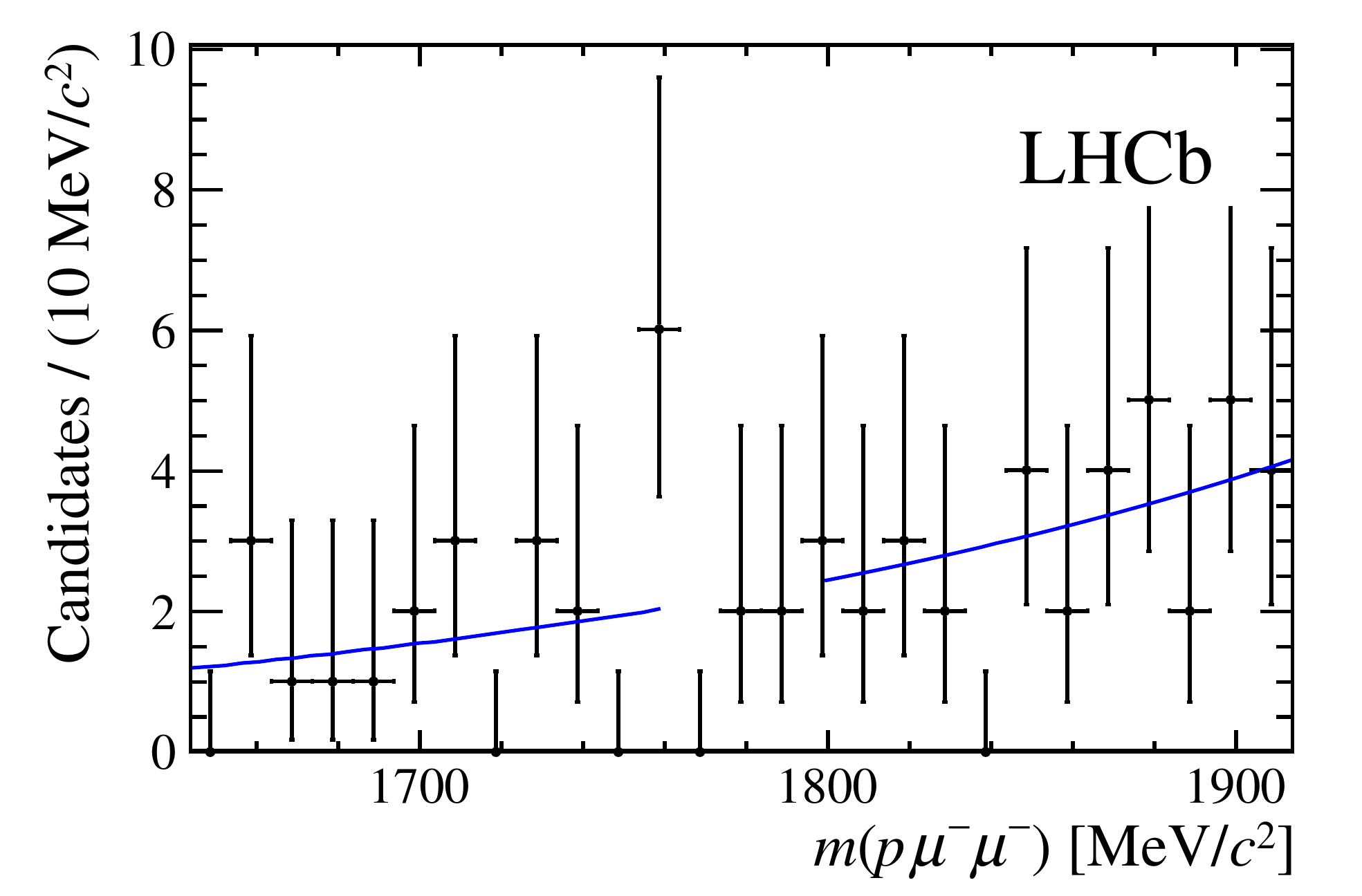}
	\put (40,120) {\small{(c)}}
	\put (35,105) {\tiny{\gl$\in [0.40, 1.0]$}}
	\end{overpic}
\end{minipage}
\caption
{\small {Invariant mass distributions and fits to the mass sidebands in data for (a) $\mu^+\mu^-\mu^-$ candidates in the four merged bins that contain 
 the highest signal probabilities, (b) ${\bar p}\mu^+\mu^-$ candidates in the two merged bins with the highest
signal probabilities, and (c)  $p\mu^-\mu^-$ candidates in the two merged bins with the highest
signal probabilities.}}
  \label{fig:bkg_fits}
\end{center}
\end{figure}
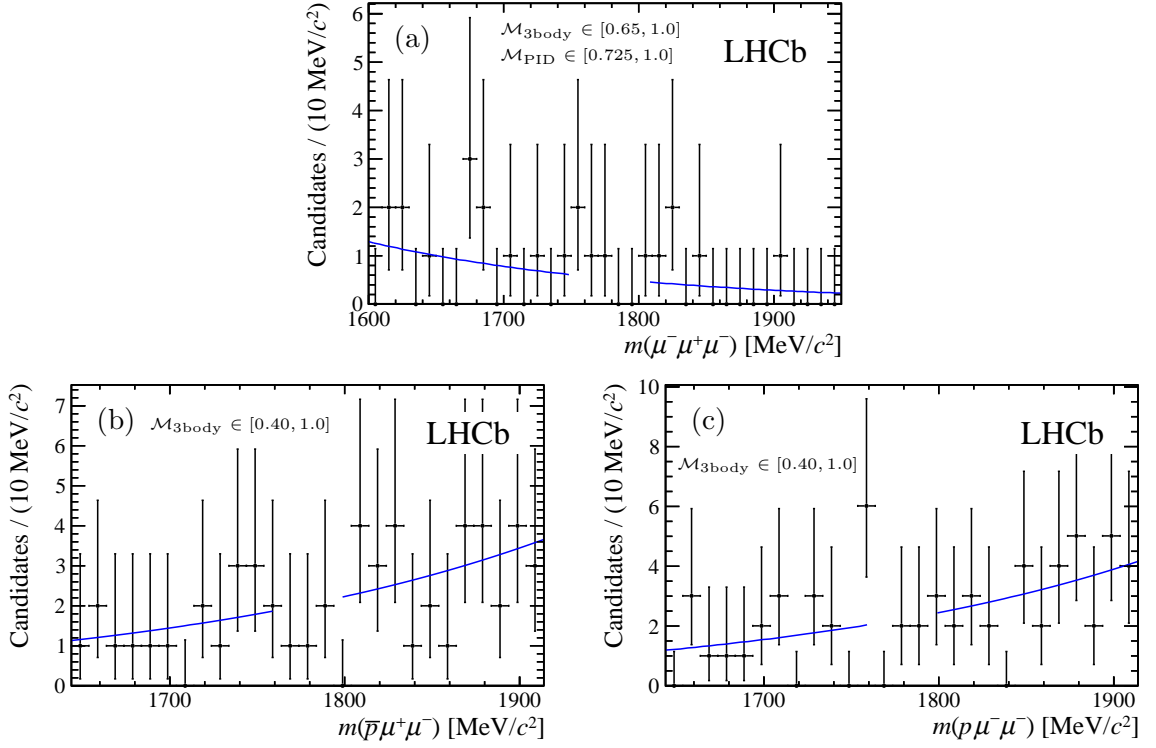

The expected numbers of background events within the signal region, for each bin in 
\gl, \pid (for \tmmm) and mass,
are evaluated by fitting the candidate mass spectra outside of the signal windows to
an exponential function using an extended, unbinned maximum likelihood fit. 
The small differences obtained if the exponential curves are 
replaced by straight lines are included as systematic uncertainties.
For \tmmm the data are fitted over the
mass range $1600 - 1950$\mevcc, while for \tpmm the fitted
mass range is $1650 - 1900$\mevcc, excluding windows around the expected signal mass of $\pm 30$\mevcc for $\mu^-\mu^+\mu^-$
and $\pm 20$\mevcc for $p\mu\mu$. 
The resulting fits to the data sidebands for a selection of bins for the three channels are shown in Fig.~\ref{fig:bkg_fits}.  

%% file: results.tex
\section{Results}
\label{sec:results}

\begin{figure}[h]
\begin{center} 
\begin{minipage}[b]{0.48\linewidth}
	\centering
	\begin{overpic}[width=\textwidth]{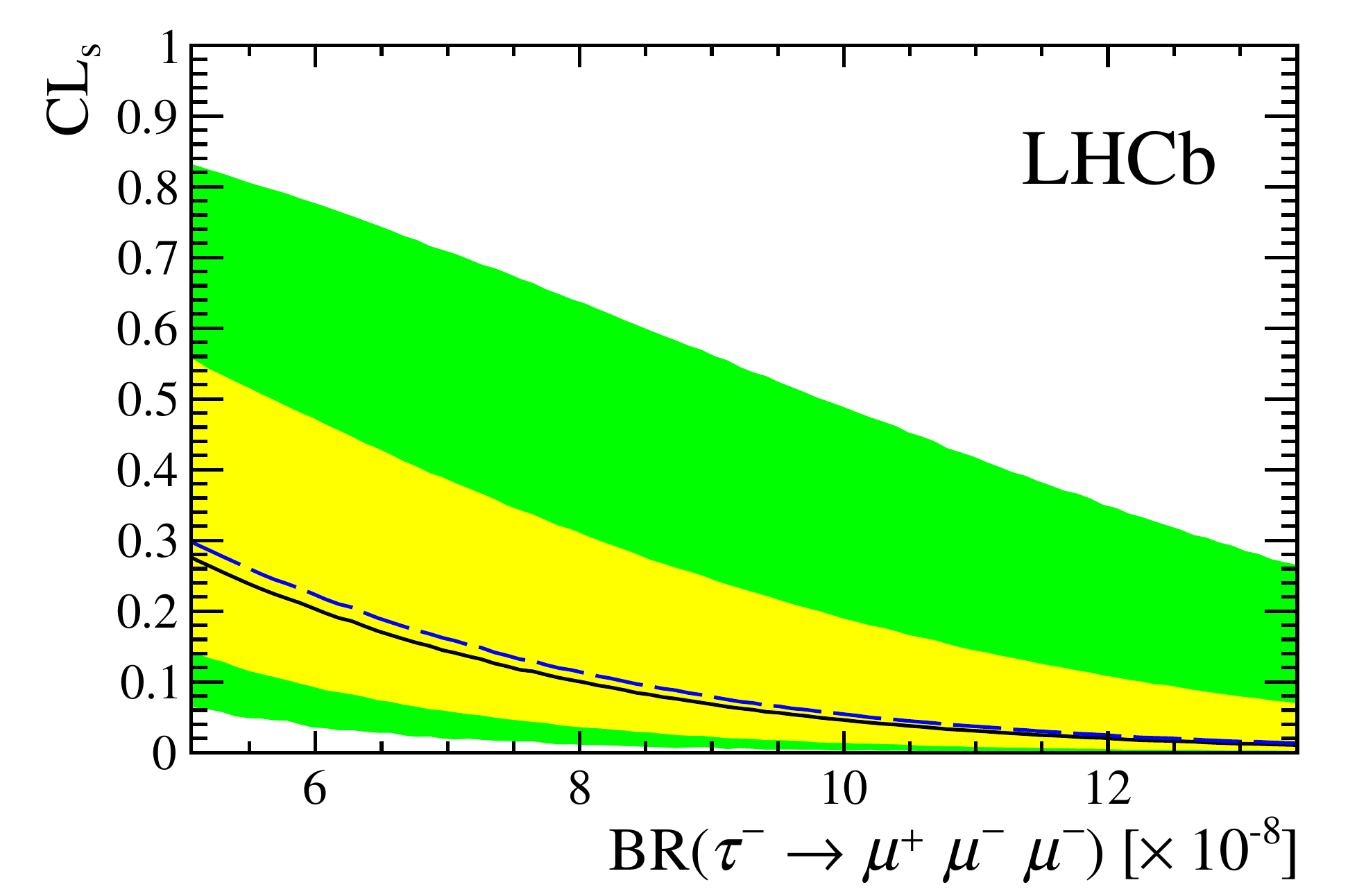}
	\put (40,120) {\small{(a)}}
	\end{overpic}
\end{minipage}\\
\begin{minipage}[b]{0.48\linewidth}
	\centering
	\begin{overpic}[width=\textwidth]{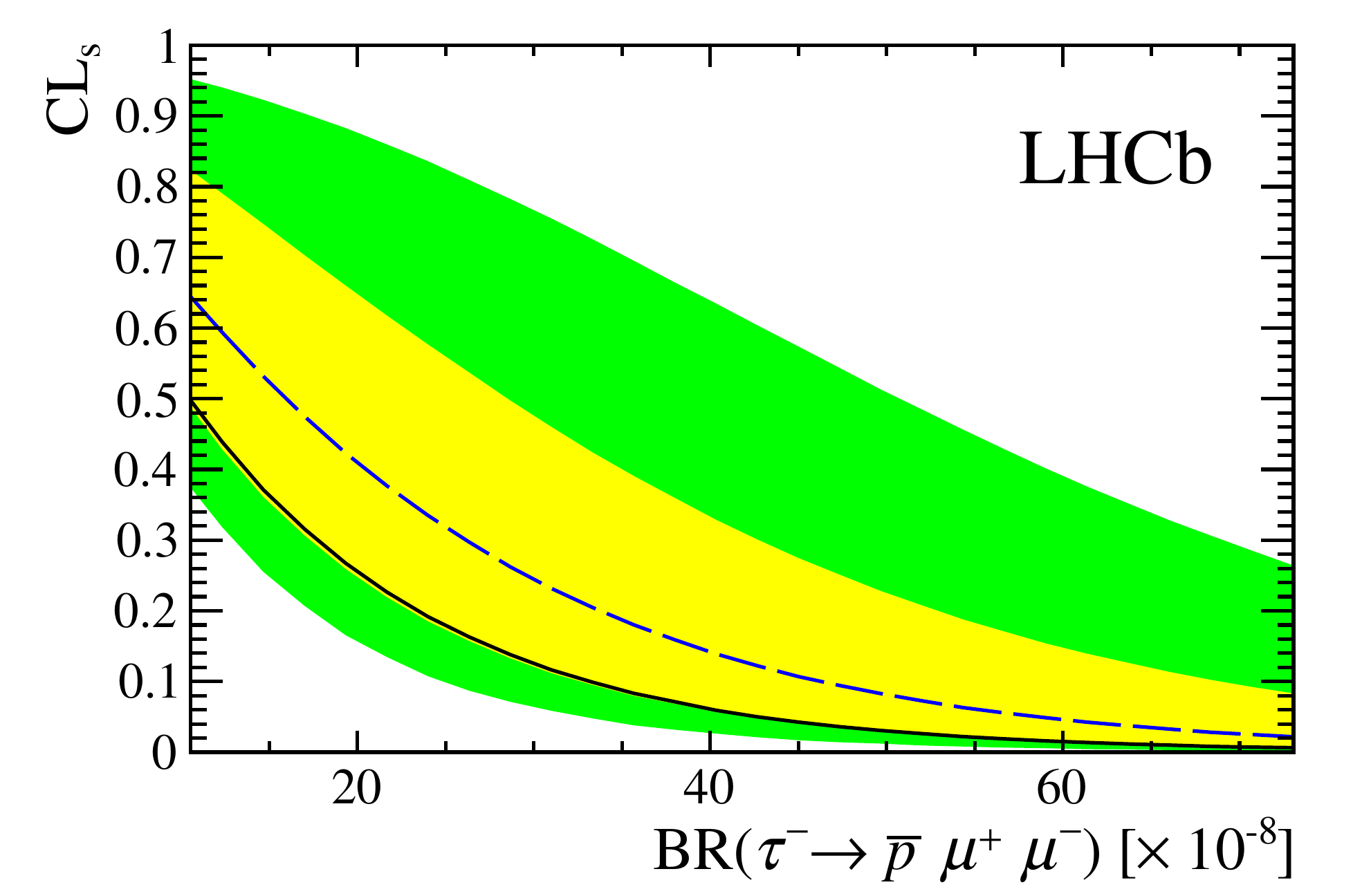}
	\put (40,120) {\small{(b)}}
	\end{overpic}
\end{minipage}
\begin{minipage}[b]{0.48\linewidth}
	\centering
	\begin{overpic}[width=\textwidth]{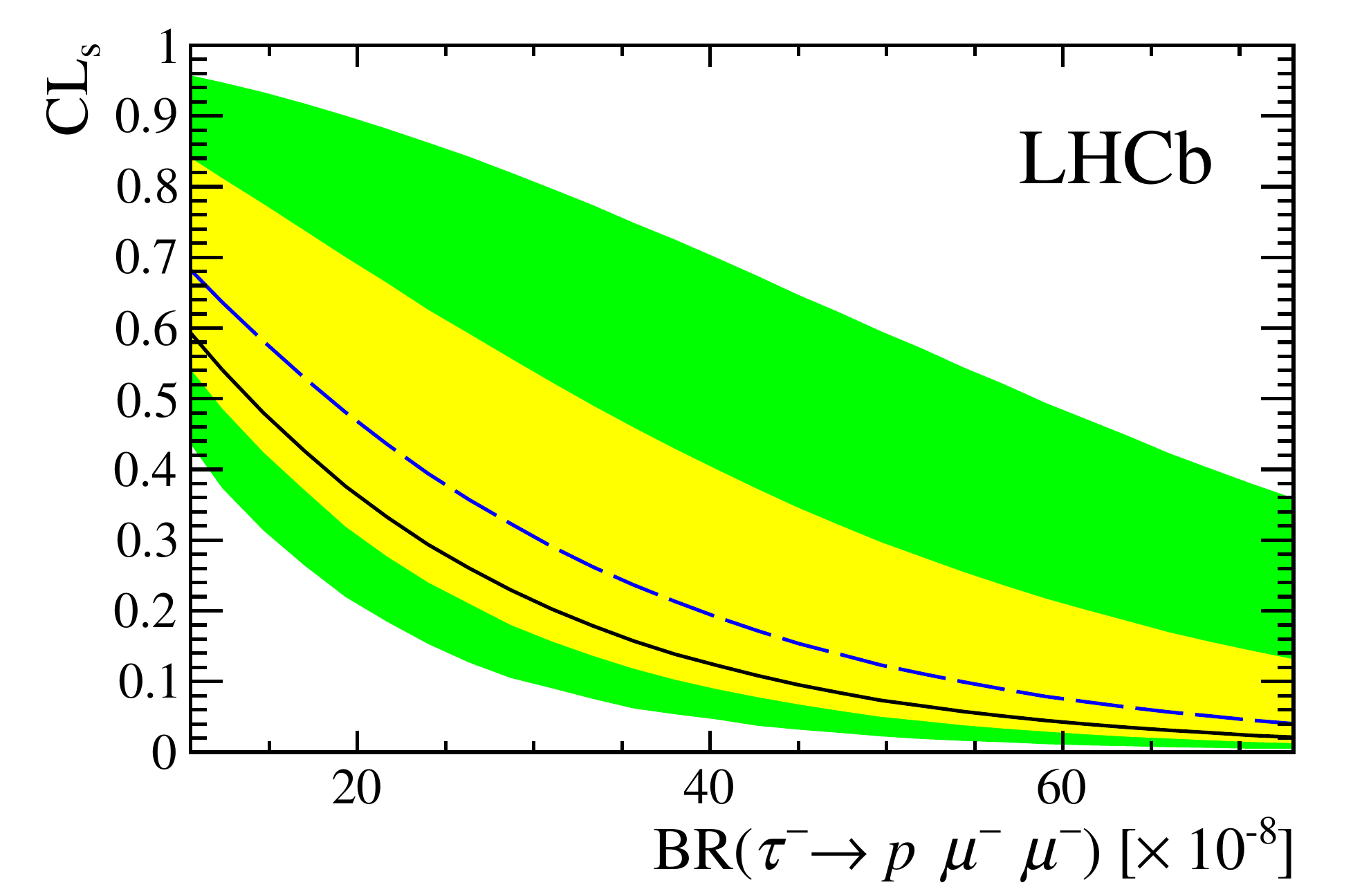}
	\put (40,120) {\small{(c)}}
	\end{overpic}
\end{minipage}
\caption
{\small {Distribution of \CLs values as functions of the assumed
  branching fractions, under the hypothesis to observe background events
  only, for (a) \tmmm, (b) \tpmmOS and (c) \tpmmSS. 
The dashed lines indicate the expected curves and the
  solid lines the observed ones.
  The light (yellow) and dark (green) bands cover the regions of 68\% and 95\%
  confidence for the expected limits.}
}
\label{fig:CLs}
\end{center} 
\end{figure}

Tables~\ref{tab:bkg_tmmm} and~\ref{tab:bkg_tpmm} give the expected and observed numbers of candidates for all
three channels investigated, in each bin of the likelihood variables, where the uncertainties on 
the background likelihoods are used to compute the uncertainties 
on the expected numbers of events. No significant evidence for an excess of events is observed.
Using the \CLs method as a statistical framework,
the distributions of observed and expected \CLs values are calculated as functions of the assumed branching fractions. 
The aforementioned uncertainties and the uncertainties on the signal likelihoods 
and normalisation factors are included using the techniques described in Ref.~\cite{Read_02, *Junk_99}.
The resulting distributions of \CLs values are shown in Fig.~\ref{fig:CLs}.

The expected limits at $90\%~(95\%)$ CL for the branching
fractions are
\begin{eqnarray*}
\BR (\tmmm) &<& 8.3~(10.2) \times 10^{-8}, \\
\BR (\tpmmOS) &<& 4.6~(5.9) \times 10^{-7},\\
\BR (\tpmmSS) &<& 5.4~(6.9) \times 10^{-7},
\end{eqnarray*}
while the observed limits at $90\%~(95\%)$ CL are
\begin{eqnarray*}
\BR (\tmmm) &<& 8.0~(9.8) \times 10^{-8},\\
\BR (\tpmmOS) &<& 3.3~(4.3) \times 10^{-7},\\
\BR (\tpmmSS) &<& 4.4~(5.7) \times 10^{-7}.
\end{eqnarray*}

All limits are given for the phase-space model of $\tau$ decays.
For \tmmm, the efficiency is found to vary by no more than $20\%$ over the
$\mu^-\mu^-$ mass range and by  $10\%$ over the $\mu^+\mu^-$ mass
range. For \tpmm, the efficiency varies by less than $20\%$ over the dimuon
mass range and less than $10\%$ with $p\mu$ mass.

In summary, a first limit on the lepton flavour violating decay mode
\tmmm has been obtained at a hadron collider.
The result is compatible with previous limits and indicates that with the
additional luminosity expected from the LHC over the coming years, the
sensitivity of LHCb will become comparable with, or exceed, those of
BaBar and Belle. First direct upper limits have been placed on the 
branching fractions for two $\tau$ decay modes that violate both baryon
number and lepton flavour, \tpmmOS and \tpmmSS.
\begin{table}[h]
\begin{center}
\caption{\small {Expected background candidate yields, with their systematic uncertainties, and observed candidate yields 
within the $\tau$ signal window in the different likelihood bins for the 
\tmmm analysis.  The likelihood values for \pid range from
$0$ (most background-like) to $+1$ (most signal-like), while
those for \gl range 
from $-1$ (most background-like) to $+1$ (most signal-like).
The lowest likelihood
bins have been excluded from the analysis. }}
\small
\label{tab:bkg_tmmm}
\begin{tabular}{c|r@{\hspace{1mm}--\hspace{1mm}}l|r@{\hspace{1mm}$\pm$\hspace{1mm}}l|c}
\pid & \multicolumn{2}{c|}{\gl} & \multicolumn{2}{c|}{Expected} & Observed\\
\hline
 & $-$0.48 & 0.05 & 345.0 & 6.7 & 409\\
 & 0.05 & 0.35 & 83.8 & 3.3 & 68\\
0.43 -- 0.6 & 0.35 & 0.65 & 30.2 & 2.0 & 35\\
 & 0.65 & 0.74 & 4.3 & 0.8 & 2\\
 & 0.74 & 1.0 & 1.4 & 0.4 & 1\\
\hline
 & $-$0.48 & 0.05 & 73.1 & 3.1 & 64\\
  & 0.05 & 0.35 & 18.3 & 1.5 & 15\\
0.6 -- 0.65 & 0.35 & 0.65 & 8.6 & 1.1 & 7\\
  & 0.65 & 0.74 & 0.4 & 0.1 & 0\\
  & 0.74 & 1.0 & 0.6 & 0.2 & 2\\
\hline
  & $-$0.48 & 0.05 & 45.4 & 2.4 & 51\\
  & 0.05 & 0.35 & 11.7 & 1.2 & 6\\
0.65 -- 0.725 & 0.35 & 0.65 & 5.3 & 0.8 & 3\\
  & 0.65 & 0.74 & 0.8 & 0.2 & 1\\
  & 0.74 & 1.0 & 0.4 & 0.1 & 0\\
\hline
  & $-$0.48 & 0.05 & 44.5 & 2.4 & 62\\
  & 0.05 & 0.35 & 10.6 & 1.2 & 13\\
0.725 -- 0.86 & 0.35 & 0.65 & 7.3 & 1.0 & 7\\
  & 0.65 & 0.74 & 1.0 & 0.2 & 2\\
  & 0.74 & 1.0 & 0.4 & 0.1 & 0\\
\hline
  & $-$0.48 & 0.05 & 5.9 & 0.9 & 7\\
  & 0.05 & 0.35 & 0.7 & 0.2 & 1\\
0.86 -- 1.0 & 0.35 & 0.65 & 1.0 & 0.2 & 1\\
  & 0.65 & 0.74 & 0.5 & 0.0 & 0\\
  & 0.74 & 1.0 & 0.4 & 0.1 & 0\\
\end{tabular} 
\end{center}
\end{table}
\begin{table}[h]
\begin{center}
\caption{\small {Expected background candidate yields, with their systematic uncertainties, and 
  observed candidate yields within the $\tau$ mass window in the different likelihood bins for the \tpmm analysis. The likelihood values
for \gl range 
from $-1$ (most background-like) to $+1$ (most signal-like). The lowest likelihood
bin has been excluded from the analysis.}}
\small
\label{tab:bkg_tpmm}
\begin{tabular}{r@{\hspace{1mm}--\hspace{1mm}}l|r@{\hspace{1mm}$\pm$\hspace{1mm}}l|c|r@{\hspace{1mm}$\pm$\hspace{1mm}}l|c}
\multicolumn{2}{c|}{} & \multicolumn{3}{|c|}{\tpmmOS} & \multicolumn{3}{c}{\tpmmSS}\\
\hline
\multicolumn{2}{c|}{\gl} & \multicolumn{2}{c|}{Expected} & Observed & \multicolumn{2}{c|}{Expected} & Observed\\
\hline
$-$0.05 & 0.20 &  37.9 & 0.8 & 43 & 41.0 & 0.9 & 41 \\\hline
0.20 & 0.40 & 12.6 & 0.5 & 8 & 11.0 & 0.5 & 13 \\\hline
0.40 & 0.70 & 6.76 & 0.37 & 6 & 7.64 & 0.39 & 10 \\\hline
0.70 & 1.00 & 0.96 & 0.14 & 0 & 0.49 & 0.12 & 0 \\
\end{tabular}
\end{center}
\end{table}

\clearpage

%% file: acknowledgements.tex
\section*{Acknowledgements}

\noindent We express our gratitude to our colleagues in the CERN
accelerator departments for the excellent performance of the LHC. We
thank the technical and administrative staff at the LHCb
institutes. We acknowledge support from CERN and from the national
agencies: CAPES, CNPq, FAPERJ and FINEP (Brazil); NSFC (China);
CNRS/IN2P3 and Region Auvergne (France); BMBF, DFG, HGF and MPG
(Germany); SFI (Ireland); INFN (Italy); FOM and NWO (The Netherlands);
SCSR (Poland); ANCS/IFA (Romania); MinES, Rosatom, RFBR and NRC
``Kurchatov Institute'' (Russia); MinECo, XuntaGal and GENCAT (Spain);
SNSF and SER (Switzerland); NAS Ukraine (Ukraine); STFC (United
Kingdom); NSF (USA). We also acknowledge the support received from the
ERC under FP7. The Tier1 computing centres are supported by IN2P3
(France), KIT and BMBF (Germany), INFN (Italy), NWO and SURF (The
Netherlands), PIC (Spain), GridPP (United Kingdom). We are thankful
for the computing resources put at our disposal by Yandex LLC
(Russia), as well as to the communities behind the multiple open
source software packages that we depend on.